\documentclass{elsart3}

\usepackage{epsfig}

\usepackage{amssymb}

\begin{document}

\begin{frontmatter}



\title{New Magnetic-Field-Induced Macroscopic Quantum Phenomenon
in High-$T_c$ Cuprates: Confined Field-Induced Density Waves in 
the Superconducting State}


\author[label1]{G. Varelogiannis} \author[label2]{M. H\'{e}ritier}

\address[label1]{Department. of Physics, National Technical University of
Athens, GR-15780 Athens, Greece}

\address[label2]{Solid State Physics Laboratory, University Paris-Sud and CNRS,
F-91405 Orsay Cedex, France}

\begin{abstract}
We reveal a novel macroscopic quantum phenomenon induced by a magnetic
field. It corresponds to the {\it non-integer quantization of the superfluid
density} in a superconductor with gap nodes due to the generation of
confined field-induced density waves (CFIDW) over a portion of the
Fermi surface (FS). The Landau numbers $L$ are not sufficient to index these
macroscopic quantum states and new quanrum numbers $\zeta$ must be added.
Distinct qualitative implications of this $|L,\zeta >$ quantization are
evident in a number of puzzling experiments in high-$T_c$ cuprates including
the plateaus behavior in the field profile of thermal conductivity,
field induced magnetic moments,
charge textures around the vortices,
and field induced vortex-solid to cascade vortex glass transitions.
\end{abstract}

\end{frontmatter}

Neutron scattering reported
the generation of AFM moments in cuprates 
inside the SC state by a magnetic field applied
perpendicular to the planes \cite{Lake}.
STM reported a field induced
checkerboard structure
that covers a region around each vortex \cite{Hoffman}.
Heat capacity and magnetization 
by Bouquet et al. \cite{Bouquet} not only
confirm the puzzling    
first order transition from a vortex
lattice to a so called vortex glass state, 
but at higher fields they
observe
a surprising transition from the vortex glass state to a new vortex glass
state which has not found any theoretical explanation so far.
In an earlier thermal transport experiment,
Krishana et al. \cite{Krishana}
have reported that,
with 
a magnetic field 
perpendicular to the $CuO_2$ planes,
the thermal conductivity shows sharp first order transitions from
a field dependent regime characteristic
of gap nodes, to a field independent regime indicating
the elimination of the nodes.
Penetration
depth studies in the presence of a stronger
than usually magnetic field 
confirm the elimination of the nodes  
adding a fundamental new element:
{\it The elimination of the nodes
is accompanied by a substantial
reduction of the superfluid density} \cite{Sonier}.

Our physical picture 
accounts  for
{\it all} these seemingly unrelated puzzling results \cite{Preprint}.
We reveal a new field induced phase transition
from a $d_{x^2-y^2}$ SC state to a
state in which $d_{x^2-y^2}$ SC {\it coexists with
Confined Field Induced (Spin and Charge) Density Waves (CFIDW) which
develop over a portion of the Fermi surface (FS) in the gap node regions}.
\begin{figure}[h]
\centerline{\psfig{figure=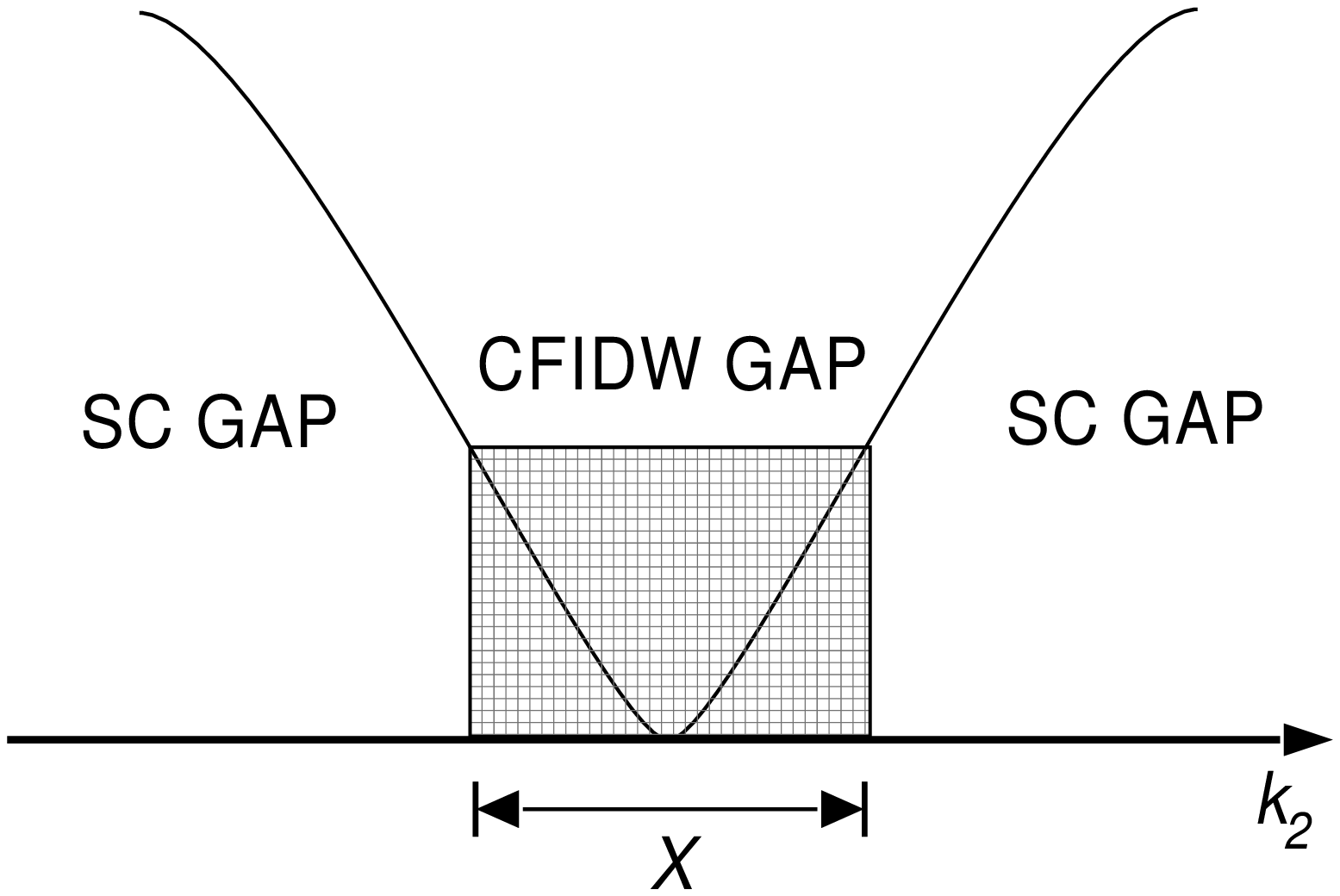,width=7.5cm,height=3.0cm,angle=0}}
\centerline{\psfig{figure=figu1b.eps,width=7.5cm,height=4.0cm,angle=0}}
\vspace{0.3cm}
\caption{
a):
Schematic view of the SC-CFIDW competition.
b):
Graphic solution of
(2) in the $L=1$ configuration and a field of 3 Tesla.
Two different relative extensions Z
($X\approx k_F \sin (Z\pi/2)$) of the CFIDW are
possible.
}
\end{figure}
Our HTSC system is built of two subsystems:
subsystem I is the Fermi Surface (FS) region covered by
the superconducting gap and subsystem II is a virtual
normal quasiparticle region created by the magnetic field
and centered in the node
points of the FS.
The
CFIDW will develop in region II because of the
orbital effect of the field in this region.
The orbital effect of the field in region I
induces vortices that are assumed to be irrelevant.

In HTSC the SC gap is $d_{x^2-y^2}$ with nodes in the
$(\pm\pi,\pm\pi)$ directions where region II is centered.
In region II {\it we necessarily have open FS sheets}.
Therefore, we can
write the dispersion of subsystem II
in the form:
$
\xi^{II}_{\bf k}=\upsilon_F(|k_1|-k_F)-2t_2\cos (k_2/X)-
2t'_2\cos (2k_2/X)
$
where $X$ is the {\it unknown} momentum extension of region II
(see Fig. 1a).
$ k_1$ is along the
$(\pm\pi,\pm\pi)$ directions perpendicular to the open FS sheets of
subsystem II,
and $k_2$ is perpendicular to $k_1$ and therefore along
the open FS sheets where we keep only two harmonics
without influence on the generality of the results.
A first order
field induced density wave gap in region II
is given by
$
\Delta_{DW}=W\exp \bigl\{ - [g N(E_F)I^2_L(X)]^{-1}\bigr\}
$
where
$$
I_L(X)=\sum_{n}J_{L-2n}\biggl({4t_2 X\over eH\upsilon_F}\biggr)
J_n\biggl({2t'_2 X\over eH\upsilon_F}\biggr)
\eqno(1)
$$
$J_n(x)$ are Bessel functions, $L$ is the index of
the Landau level configuration,
$e$ is the charge of the electron, $H$ the magnetic field,
$N(E_F)$ the DOS at the Fermi level (in region II), $g$
a scattering amplitude (Coulombic or phononic),
$W$ the bandwidth in the $(\pi,\pi)$ direction
and $\upsilon_F$ the
Fermi velocity.

CFIDW states will develop only if
the absolute free energy gain 
due to the opening of a CFIDW gap in
region II is bigger than that
lost by the elimination of the SC gap
from this region.
Moreover,
the CFIDW state must be {\it confined in momentum space}
with a DW
gap smaller or equal to the absolute SC gap in the
borders of region II 
$$
I^2_L\bigl(k_F\sin(Z\pi/2)\bigr)\leq
{1 \over g N(E_F)\ln {W\over \Delta_{sc}\sin(Z\pi/2)}}
\eqno(2)
$$
the equality fixing the
relative extension $Z$ 
of the CFIDW for each $L$ ($X\approx k_F \sin (Z\pi/2)$)
and therefore
$I^2_L(Z)$ which then
fixes the CFIDW gap $\Delta_{DW}$
and the critical temperature $T_{DW}$ at which the CFIDW forms.
$\Delta_{sc}$ is the maximum value of the $d_{x^2-y^2}$ SC gap.
A graphic solution of (2) for Z is shown in Fig. 1b.
Surprisingly, there are two possible values of Z for a given Landau level
configuration $L$. A {\it new quantum number} $\zeta$
must be added to the Landau numbers $L$, which will index 
{\it the consequently non-integer quantization of the superfluid
density}. Indeed, each quantized value of Z corresponds to a different
relative extension of the SC region over the FS
and therefore to a different density of superfluid carriers.
Field induced cascade transitions between $|L,\zeta>$ configurations
explain the cascade vortex-glass transitions of \cite{Bouquet}
indicating that vortex glass states are associated with our
SC+CFIDW states. For details see Ref. \cite{Preprint}.

\end{document}